\documentclass[conference]{IEEEtran}
\usepackage[utf8]{inputenc}
\usepackage{mathptmx}
\usepackage{amssymb}
\usepackage{amsmath}
\usepackage{cite}
\usepackage{graphicx}
\usepackage{tikz}

\newtheorem{remark}{\bf Remark}[section]

\newtheorem{example}{\bf Example}[section]

\ifCLASSINFOpdf
\else
\fi
\begin{document}
%
\title{MDP based Decision Support for Earthquake Damaged Distribution System Restoration}



%
\author{
\IEEEauthorblockN{Onur Yigit Arpalı\IEEEauthorrefmark{1},
Ugur Can Yilmaz\IEEEauthorrefmark{2}, 
Ebru Aydin Gol\IEEEauthorrefmark{1}
Burcu Guldur Erkal\IEEEauthorrefmark{3} and
Murat Gol\IEEEauthorrefmark{2}}%
\IEEEauthorblockA{\IEEEauthorrefmark{1}Dept. of Computer Eng.,
METU,Ankara, Turkey\\ Email: arpali.onur@metu.edu.tr , ebrugol@metu.edu.tr }
\IEEEauthorblockA{\IEEEauthorrefmark{2}Dept. of Electrical \& Electronics Eng.,
METU,Ankara, Turkey\\ Email: ugur.yilmaz@metu.edu.tr , mgol@metu.edu.tr }
\IEEEauthorblockA{\IEEEauthorrefmark{3}Dept. of Civil Eng.,
Hacettepe University,Ankara, Turkey\\ Email: burcuguldur@hacettepe.edu.tr }%
}


\maketitle

\begin{abstract}
As the society becomes more dependent on the presence of electricity, the resilience of the power systems gains more importance. This paper develops a decision support method for distribution system operators to restore electricity after an earthquake to the maximum number of customers in the minimum expected duration. The proposed method employs Markov Decision Process (MDP) to determine the optimal restoration scheme. In order to determine the probability of the field component damage due to the earthquake, the Probability of Failure ($P_f$) of  structures are calculated using the Peak Ground Acceleration (PGA) values recorded by observatories and earthquake research centers during the earthquake. \end{abstract}



%
\IEEEpeerreviewmaketitle

\section{Introduction}

Restoring electricity\footnote{$ \textcopyright$ 2019 IEEE.  Personal use of this material is permitted.  Permission from IEEE must be obtained for all other uses, in any current or future media, including reprinting/republishing this material for advertising or promotional purposes, creating new collective works, for resale or redistribution to servers or lists, or reuse of any copyrighted component of this work in other works. \textit{Under review.}} fast after an earthquake is extremely important, as the modern life relies on the presence of electricity. After an earthquake, some of the field instruments may get damaged.  Considering that black-start of a system without any structural damage is a difficult problem, restoration at a damaged system has many challenges. This paper proposes a decision support method for distribution systems with damaged instruments due to an earthquake. The proposed method uses the probability of exceeding the damage state of failure (Probability of Failure - $P_f$) to determine the best restoration strategy. Central to the proposed method is the construction of a Markov Decision Process that encapsulates the system topology, electrical constraints and damage probabilities ($P_f$).  

There are studies in the literature on the power system restoration and disaster management problems \cite{Arab2015, Qiu2017, Yuan2016, Wang2019, Zhao2018, Golshani2017, Ganganath2018, FerreiraNeto2016, Ostermann2017 }.  While \cite{Arab2015} and \cite{Qiu2017} develop recovery plans for expected disasters, \cite{Yuan2016} and \cite{Wang2019} presents planning strategies to improve the power system resilience. In \cite{Zhao2018},  use of micro-grids in fast restoration of power systems is evaluated. While \cite{Golshani2017, Ganganath2018, FerreiraNeto2016, Ostermann2017} present online disaster management methods based on field sensor data, in those studies field component damage due to the earthquake is neglected. 

This paper proposes an MDP based decision support method to restore a medium voltage (MV) distribution system after an earthquake. The method generates a restoration strategy that minimizes the average expected restoration time over the buses considering the real time data of the earthquake. The proposed decision support method does not require any additional infrastructure, such that basic SCADA data and $P_f$, which is determined from observatories, are sufficient to run the method.

The framework of the proposed method was first introduced in \cite{AydinGol2019}. This paper enhances the model presented in \cite{AydinGol2019} by considering electrical constraints, such that a power flow algorithm runs along with the proposed method in order to check if voltages of the system remain within the accepted limits during the restoration actions. Moreover, Distributed Energy Resources (DERs) are also included in the problem solution. In~\cite{AydinGol2019}, the objective was defined as minimizing the total restoration time. However, it does not necessarily minimize the expected restoration time of the buses. This objective is achieved with the updated cost formulation. Note that, the proposed method is designed to run after the earthquake occurs. The method utilizes real time earthquake data, and determines the restoration strategy considering the field component damage probabilities.

The paper is organized as follows. The background information is given in Section II and the proposed method is defined in detail in Section III. Numerical validation is provided in Section IV, followed by conclusions in Section V.


\section{Preliminaries}

\subsection{Probability of Failure ($P_f$)}\label{sec:PF}
Structures are constructed to resist the adverse effects of reoccurring natural events such as earthquakes. Current design specifications ensure the safety of a structure after a major event with a certain return period. However, some of the existing structures do not conform to the criteria specified by the current design specifications. The behavior of these critical structures during and after a major event could be estimated by conducting a fragility analysis.

Fragility analysis is performed to estimate the seismic loss of built environments. The obtained fragility curves represent the probability of exceeding a damage limit state for a given structure subjected to seismic excitation. For this research, the critical limit state is defined as collapse, and the fragility curves obtained for the investigated structures, therefore, consist of peak ground acceleration (PGA) versus $P_f$ values. Thus, by using the obtained fragility curves, it is possible to get the $P_f$ values associated with each structure for particular excitation values (PGA). Note that PGA values are calculated using the earthquake data recorded by the observatories. 

\subsection{Markov Decision Process (MDP)}
MDP is a mathematical framework that is used to make decisions in probabilistic environments.
The policy synthesis problem for MDPs concerns the generation of a policy optimising the expected cost under the given constraints~\cite{Bertsekas:2007:DPO:1396348}.

A Markov Decision Process (MDP) is defined by the tuple $M = (S, A, p, c)$, 
where $S$ is a set of states, $A$ is a set of actions, $p : S \times A \times S \rightarrow
[0,1]$ is a probabilistic action-conditioned state transition function, i.e.,  
given a state $s\in S$ and action $a \in A$, $p(s^{'}|s,a)$ is the probability  of 
transitioning from $s$ to $s^{'}$ when action $a$ is applied, and $c:S \rightarrow \mathbb{R}_+$
is a state cost function. The set of actions that can be applied in a state $s$ is denoted by
$A(s)$. A deterministic policy $\pi$ gives the action to be applied in a state $s$ $\pi : S \rightarrow A$ with $\pi(s) \in A(s)$. 
A value function $v_{\pi}^n : S \rightarrow \mathbb{R}$ represents the expected cost obtained when policy $\pi$ is followed for $n$ steps and it is recursively defined as follows:
\begin{equation} \label{eq:erl}
    v_{\pi}^n(s) = \begin{cases} c(s) & \text{ if }  n = 1 \\
    c(s) + \sum_{s^{'} \in S} p(s^{'}|s,\pi(s))  v_{\pi}^{n-1}(s^{'}) &  \text{ otherwise }
    \end{cases}
\end{equation}



For a given state $s_0 \in S$, the optimal policy $\pi^{*}$ is the policy
minimizing the value function, $\pi^{*} = \underset{\pi}{argmin}$ $v^n_{\pi}(s_0)$.

\subsection{Power Flow Analysis}

\par Among the power flow analysis methods, such as Newton-Raphson (NR), Gauss-Seidel (GS), fast decoupled load flow (FDLF), forward-backward power flow (FBPF)\cite{Grainger1994,Ghosh1999}, the proposed method utilizes FBPF method, because of the radial structure distribution networks and high R/X ratio of the distribution system power lines \cite{Muruganantham2017}. Implementation details of FBPF can be found in \cite{Ghosh1999}. The paper assumes proper load and generation forecasts are provided, as the paper does not develop any forecast algorithm.

\par This work assumes that DERs do not contribute to voltage control, such that they aim to supply only active power, which is a realistic assumption considering operating principles of renewable sources. Therefore, DERs are modeled as negative power demands in the FBPF algorithm. Note that, a DER will only contribute to voltage control if it operates at island mode.

Considering that the system will be interrupted, to simplify the problem, the FBPF solves single-phase equivalent network, such that the system is assumed to be balanced. Once the system is restored, the loads may be unbalanced, however, as the medium voltage system is considered, the unbalance would be limited.

\section{Proposed Method}
In this section, the proposed MDP based restoration policy synthesis approach for the distribution system is explained. 
Throughout the section, the number of buses, branches, and DERs of the considered distribution system are represented by $N$, $L$, and $M$, respectively. The notation $I_K$ is used to denote the set of positive integers less than or equal to $K$, i.e., $I_K = \{1,\ldots,K\}$


\subsection{Model Construction}
In the proposed MDP model $M = (S, A, p, c)$, each state $s \in S$ represents the current situation of
all branches of the system. Thus, each state is a snapshot of the distribution system. A branch can be
(1) \textit{damaged} ($D$), or it has not been tried to energized yet, thus its health condition can be
(2) \textit{unknown} ($U$), or it can be \textit{energized} ($E_i$), where the subscript $i \in \{0\} 
\cup I_L$ denotes the major source of its energy. The index is $0$ when the source is the transmission grid,
when $i\in I_L$, it is the index of the DER feeding the branch. Thus, the set of states $S$ is:
\begin{equation}\label{eq:states}
\begin{gathered} 
    S = \{s_0, s_1, \dots, s_F\} 
    \text{ where }\ s_i = [s_i^1, s_i^1, \dots, s_i^{L}] \ \text{ and }\\
    s_i^k \in \{U,D\} \cup \{E_0,E_1, \ldots, E_M\}.
\end{gathered}
\end{equation}

\begin{figure}[ht]
\vspace{-6pt}
\centering
\tikzset{every picture/.style={line width=0.75pt}} 

\begin{tikzpicture}[x=0.75pt,y=0.75pt,yscale=-1,xscale=1]

\draw  [fill={rgb, 255:red, 0; green, 0; blue, 0 }  ,fill opacity=1 ][line width=0.75]  (182.24,139.84) .. controls (183.76,139.84) and (185,141.07) .. (185,142.6) -- (185,177.24) .. controls (185,178.76) and (183.76,180) .. (182.24,180) -- (182.24,180) .. controls (180.71,180) and (179.48,178.76) .. (179.48,177.24) -- (179.48,142.6) .. controls (179.48,141.07) and (180.71,139.84) .. (182.24,139.84) -- cycle ;
\draw    (394.39,191.02) -- (462.57,191.24) ;

\draw  [fill={rgb, 255:red, 65; green, 117; blue, 5 }  ,fill opacity=1 ] (398.94,187.45) -- (406.61,187.45) -- (406.61,195.12) -- (398.94,195.12) -- cycle ;
\draw  [fill={rgb, 255:red, 65; green, 117; blue, 5 }  ,fill opacity=1 ] (451.48,187.49) -- (459.14,187.49) -- (459.14,195.15) -- (451.48,195.15) -- cycle ;
\draw  [fill={rgb, 255:red, 0; green, 0; blue, 0 }  ,fill opacity=1 ][line width=0.75]  (256.24,139.84) .. controls (257.76,139.84) and (259,141.07) .. (259,142.6) -- (259,177.24) .. controls (259,178.76) and (257.76,180) .. (256.24,180) -- (256.24,180) .. controls (254.71,180) and (253.48,178.76) .. (253.48,177.24) -- (253.48,142.6) .. controls (253.48,141.07) and (254.71,139.84) .. (256.24,139.84) -- cycle ;
\draw  [fill={rgb, 255:red, 0; green, 0; blue, 0 }  ,fill opacity=1 ][line width=0.75]  (325.24,139.84) .. controls (326.76,139.84) and (328,141.07) .. (328,142.6) -- (328,177.24) .. controls (328,178.76) and (326.76,180) .. (325.24,180) -- (325.24,180) .. controls (323.71,180) and (322.48,178.76) .. (322.48,177.24) -- (322.48,142.6) .. controls (322.48,141.07) and (323.71,139.84) .. (325.24,139.84) -- cycle ;
\draw  [fill={rgb, 255:red, 0; green, 0; blue, 0 }  ,fill opacity=1 ][line width=0.75]  (396.24,139.84) .. controls (397.76,139.84) and (399,141.07) .. (399,142.6) -- (399,177.24) .. controls (399,178.76) and (397.76,180) .. (396.24,180) -- (396.24,180) .. controls (394.71,180) and (393.48,178.76) .. (393.48,177.24) -- (393.48,142.6) .. controls (393.48,141.07) and (394.71,139.84) .. (396.24,139.84) -- cycle ;
\draw  [fill={rgb, 255:red, 0; green, 0; blue, 0 }  ,fill opacity=1 ][line width=0.75]  (465.24,139.84) .. controls (466.76,139.84) and (468,141.07) .. (468,142.6) -- (468,177.24) .. controls (468,178.76) and (466.76,180) .. (465.24,180) -- (465.24,180) .. controls (463.71,180) and (462.48,178.76) .. (462.48,177.24) -- (462.48,142.6) .. controls (462.48,141.07) and (463.71,139.84) .. (465.24,139.84) -- cycle ;
\draw    (182.24,159.92) -- (256.24,159.92) ;

\draw    (256.24,159.92) -- (345.24,159.92) ;

\draw    (342.24,159.92) -- (395.24,159.92) ;

\draw    (392.24,159.92) -- (465.24,159.92) ;

\draw  [fill={rgb, 255:red, 0; green, 0; blue, 0 }  ,fill opacity=1 ][line width=0.75]  (325.24,185.84) .. controls (326.76,185.84) and (328,187.07) .. (328,188.6) -- (328,223.24) .. controls (328,224.76) and (326.76,226) .. (325.24,226) -- (325.24,226) .. controls (323.71,226) and (322.48,224.76) .. (322.48,223.24) -- (322.48,188.6) .. controls (322.48,187.07) and (323.71,185.84) .. (325.24,185.84) -- cycle ;
\draw    (256,170.67) -- (268,170.67) ;

\draw  [fill={rgb, 255:red, 65; green, 117; blue, 5 }  ,fill opacity=1 ] (187.38,155.29) -- (195.05,155.29) -- (195.05,162.95) -- (187.38,162.95) -- cycle ;
\draw  [fill={rgb, 255:red, 65; green, 117; blue, 5 }  ,fill opacity=1 ] (243.38,156.29) -- (251.05,156.29) -- (251.05,163.95) -- (243.38,163.95) -- cycle ;
\draw  [fill={rgb, 255:red, 65; green, 117; blue, 5 }  ,fill opacity=1 ] (263.38,156.29) -- (271.05,156.29) -- (271.05,163.95) -- (263.38,163.95) -- cycle ;
\draw  [fill={rgb, 255:red, 65; green, 117; blue, 5 }  ,fill opacity=1 ] (311.38,155.29) -- (319.05,155.29) -- (319.05,162.95) -- (311.38,162.95) -- cycle ;
\draw  [fill={rgb, 255:red, 255; green, 255; blue, 255 }  ,fill opacity=1 ] (331.38,156.29) -- (339.05,156.29) -- (339.05,163.95) -- (331.38,163.95) -- cycle ;
\draw  [fill={rgb, 255:red, 255; green, 255; blue, 255 }  ,fill opacity=1 ] (382.38,156.29) -- (390.05,156.29) -- (390.05,163.95) -- (382.38,163.95) -- cycle ;
\draw  [fill={rgb, 255:red, 65; green, 117; blue, 5 }  ,fill opacity=1 ] (401.38,156.29) -- (409.05,156.29) -- (409.05,163.95) -- (401.38,163.95) -- cycle ;
\draw  [fill={rgb, 255:red, 65; green, 117; blue, 5 }  ,fill opacity=1 ] (451.38,156.29) -- (459.05,156.29) -- (459.05,163.95) -- (451.38,163.95) -- cycle ;
\draw    (268,170.67) -- (268,206) ;

\draw    (268,206) -- (325.24,205.92) ;

\draw  [fill={rgb, 255:red, 255; green, 255; blue, 255 }  ,fill opacity=1 ] (262.38,168.29) -- (270.05,168.29) -- (270.05,175.95) -- (262.38,175.95) -- cycle ;
\draw  [fill={rgb, 255:red, 255; green, 255; blue, 255 }  ,fill opacity=1 ] (312.38,201.29) -- (320.05,201.29) -- (320.05,208.95) -- (312.38,208.95) -- cycle ;
\draw    (393.39,214.02) -- (461.57,214.24) ;

\draw  [fill={rgb, 255:red, 255; green, 255; blue, 255 }  ,fill opacity=1 ] (397.94,210.45) -- (405.61,210.45) -- (405.61,218.12) -- (397.94,218.12) -- cycle ;
\draw  [fill={rgb, 255:red, 255; green, 255; blue, 255 }  ,fill opacity=1 ] (450.48,210.49) -- (458.14,210.49) -- (458.14,218.15) -- (450.48,218.15) -- cycle ;

\draw (191.46,145.42) node  [align=left] {{\footnotesize 1}};
\draw (429.36,201.45) node [scale=0.7] [align=left] {Energized Branch};
\draw (266.8,146.09) node  [align=left] {{\footnotesize 2}};
\draw (334.8,192.09) node  [align=left] {{\footnotesize 3}};
\draw (337.13,146.75) node  [align=left] {{\footnotesize 4}};
\draw (407.13,146.09) node  [align=left] {5};
\draw (219.33,165.67) node  [align=left] {{\footnotesize Branch 1}};
\draw (295.33,168.33) node  [align=left] {{\footnotesize Branch 2}};
\draw (359.33,169) node  [align=left] {{\footnotesize Branch 4}};
\draw (430.67,169) node  [align=left] {{\footnotesize Branch 5}};
\draw (295.33,213.67) node  [align=left] {{\footnotesize Branch 3}};
\draw (477.13,149.42) node  [align=left] {6};
\draw (428.36,224.45) node [scale=0.7] [align=left] {Unenergized Branch};

\end{tikzpicture}
\vspace{-6pt}
\caption{Sample system with 5 branches}
\vspace{-6pt}
\label{fig:sample}
\end{figure}
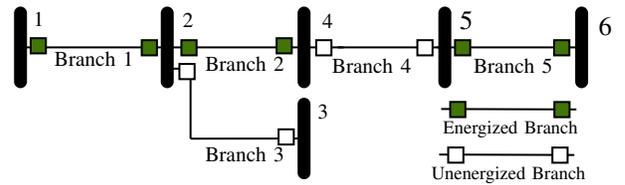

The total number of states, $F+1$, is upper bounded by $(3+M)^L$ according to~\eqref{eq:states}. However, 
most of these states represent infeasible system configurations, e.g. energized branches unconnected to a 
source, a DER supplying higher energy than its capacity, or meshed structures. Such states are never added 
to the model. Thus, in practice, the size of $S$ is much less than the given bound. Initially, all circuit
breakers are assumed to be open. Thus, the initial state of the system is $ s_0=[U,U,\dots,U]$. 
\begin{example}\label{ex:simple}
A network with 5 branches is shown in Fig \ref{fig:sample}, where node 1 is connected to the transmission grid and node 6 is connected to a DER. The corresponding MDP state is $s_2 = [E_0,E_0,U,U,E_1]$ which indicates that the first and second branches are energized from the transmission grid, the circuit breakers for the third and forth branches are open and their conditions are unknown to the system operator, and the fifth branch is energized from the first DER.
\end{example}

\subsection{Action Set Construction}
In a distribution network, a restoration action can be simultaneously applied to a set of branches that satisfy topological and electrical constraints. In the MDP model, an action $a\in A$ is defined as the set of branches to be energized simultaneously, 
\begin{equation}
\begin{gathered}
A=\{ a | a \subseteq I_L \}
\end{gathered}
\end{equation}
and the constraints are integrated to the model trough the feasible action sets $A(s)$ for each $s\in S$.

The restoration action can be applied to a branch $i$ if is in unknown condition (breakers are open) and it is connected to an energized branch or a source. Let $En \subset I_L$ denote the branches that are connected to a source and $B(i) \subset I_L$ denote the indices of branches that are physically connected to branch $i$, i.e., share a common end. Thus the set of branches for which a restoration action can be applied in a state $s=[s^1,\ldots,s^L]$ is defined as 
\begin{align}\label{eq:actionUconstraint}
\bar A (s) = \{i \mid s^i = U & \text{ and } ( i\in En \text{ or } \\
& s^j \in \{E_0,E_1,\ldots, E_M\} \text{ for some } j\in B(i) ) \} \nonumber
\end{align}

Given a state $s$, the restoration action can be simultaneously applied to branches from the set $\bar A (s) \subset I_L$ that satisfy the topological and electrical constraints. These constraints are defined with respect to the states that are one step reachable from $s$, and they are formalized following the definition of the state transition function in Sec.~\ref{sec:actionConstraints}. 

\subsection{Transition Function Calculation}
In state $s=[s^1,s^1,…s^L]$, when action $a\in A(s) \subseteq \bar A (s)$ is applied, the probability of transitioning to a state $t=[t^1,t^1,…t^L]$ satisfying~\eqref{eq:transition_condition} is given in~\eqref{eq:transition_probability}. Essentially, by $\bar A(s)$~\eqref{eq:actionUconstraint} definition $s^i=U$ for all $i \in a$ and the status of these branches change to $D$ or $E_k$ according to $P_f$ (see Section~\ref{sec:PF}), where $k$ denote the source for the energized branch that has a common end with $i$ (C3). In addition, for a branch energized from a DER, if an energized path to the transmission grid emerges as a result of performing the restoration actions to the branches in $a$, the status of this branch changes to $E_0$ (C1), as after a connection with the transmission grid is established, the capacity limitation of the DER will vanish. No other branch can change its status after $a$ is applied (C2). The probability of transitioning to a state that does not satisfy~\eqref{eq:transition_condition} is $0$ as those states are not one step reachable from $s$ under the control action $a$.
\begin{align}\label{eq:transition_condition}
    & \quad\quad\quad\quad\quad\quad t = [t^1,t^1  ,…t^L] \quad\quad \text{ where } \\
    & t^i = \begin{cases}
    E_0 \quad \text{ if }  s^i = E_k, \text{ for some } k\in I_M \text{ and there is an} \\ \text{energized path from $i$ to the transmission grid in } t \text{(C1)} \\
    s^i \quad \text{ if C1 does not hold and } i\not \in a \quad\quad\quad\quad\quad\quad\ \text{      (C2)}\\
    E_k  \text{ or } D \text{ if } i\in a \text{ and } s^j = E_k \text{ for some } j \in B(i) \quad \text{    (C3) } \nonumber
    \end{cases}
\end{align}

\begin{equation}\label{eq:transition_probability}
\begin{gathered} 
    p(t|s,a)= \prod_{i\in a}\begin{cases}
    P_f(i), & \text{if } t^i = D \\
    1-P_f(i), & \text{if } t^i = E_k, k\geq 0
  \end{cases}\\
\end{gathered}
\end{equation}
The set of states that can be reached from $s$ when the control action $a$ is applied is defined with respect to $p(\cdot \mid s,a)$:
\begin{equation}\label{eq:post}
    Post(s,a) = \{t \in S \mid p(t\mid s, a) > 0  \}
\end{equation}

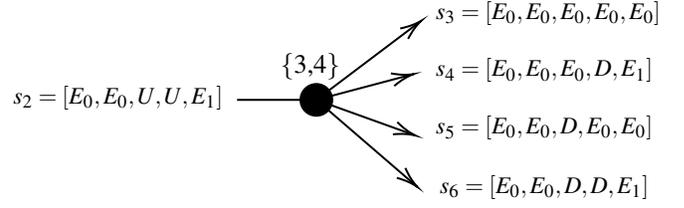
\begin{figure}[ht]
\centering

\tikzset{every picture/.style={line width=0.75pt}} 

\begin{tikzpicture}[x=0.75pt,y=0.75pt,yscale=-1,xscale=1]

\draw    (181,135) -- (213,135) ;

\draw  [fill={rgb, 255:red, 0; green, 0; blue, 0 }  ,fill opacity=1 ] (213,135) .. controls (213,130.58) and (216.58,127) .. (221,127) .. controls (225.42,127) and (229,130.58) .. (229,135) .. controls (229,139.42) and (225.42,143) .. (221,143) .. controls (216.58,143) and (213,139.42) .. (213,135) -- cycle ;
\draw    (221,135) -- (272.4,96.2) ;
\draw [shift={(274,95)}, rotate = 502.96] [color={rgb, 255:red, 0; green, 0; blue, 0 }  ][line width=0.75]    (10.93,-3.29) .. controls (6.95,-1.4) and (3.31,-0.3) .. (0,0) .. controls (3.31,0.3) and (6.95,1.4) .. (10.93,3.29)   ;

\draw    (221,135) -- (268.07,122.51) ;
\draw [shift={(270,122)}, rotate = 525.14] [color={rgb, 255:red, 0; green, 0; blue, 0 }  ][line width=0.75]    (10.93,-3.29) .. controls (6.95,-1.4) and (3.31,-0.3) .. (0,0) .. controls (3.31,0.3) and (6.95,1.4) .. (10.93,3.29)   ;

\draw    (221,135) -- (269.12,152.32) ;
\draw [shift={(271,153)}, rotate = 199.8] [color={rgb, 255:red, 0; green, 0; blue, 0 }  ][line width=0.75]    (10.93,-3.29) .. controls (6.95,-1.4) and (3.31,-0.3) .. (0,0) .. controls (3.31,0.3) and (6.95,1.4) .. (10.93,3.29)   ;

\draw    (221,135) -- (270.49,177.69) ;
\draw [shift={(272,179)}, rotate = 220.79] [color={rgb, 255:red, 0; green, 0; blue, 0 }  ][line width=0.75]    (10.93,-3.29) .. controls (6.95,-1.4) and (3.31,-0.3) .. (0,0) .. controls (3.31,0.3) and (6.95,1.4) .. (10.93,3.29)   ;

\draw (121,134) node  [align=left] {{\small $s_2 =[E_0,E_0,U,U,E_1]$}};
\draw (338,91) node  [align=left] {{\small $s_{3} =[E_0,E_0,E_0,E_0,E_0]$}};
\draw (336,149) node  [align=left] {{\small $s_{5} =[E_0,E_0,D,E_0,E_0]$}};
\draw (336,120) node  [align=left] {{\small $s_{4} =[E_0,E_0,E_0,D,E_1]$}};
\draw (336,179) node  [align=left] {{\small $s_{6} =[E_0,E_0,D,D,E_1]$}};
\draw (218,118) node  [align=left] {\{3,4\}};

\end{tikzpicture}

\caption{$Post(s_2,\{3,4\})$ for $s_2$ illustrated in Fig.~\ref{fig:sample}} \label{fig:transition}
\vspace{-12pt}
\end{figure}

\begin{example}\label{ex:transition}
The set of branches for which a restoration action can be applied is $\overline A(s_2) = \{3,4\}$ for the MDP state $s_2$ from Ex.~\ref{ex:simple}.  $Post(s_2,\{3,4\})$ is shown in Fig.~\ref{fig:transition} 
\end{example}

\subsection{Feasible Action Set Construction}\label{sec:actionConstraints}
The subsets $a $ of $\bar A (s)$~\eqref{eq:actionUconstraint} that satisfy the topological and electrical constraints are added to $A(s)$. These constraints are formalized in this section.


\noindent\textbf{Distance constraint:} The restoration action can be applied to two branches simultaneously if they are electrically distant enough such that the transients emerging after the closure of the circuit breakers do not cause significant effects. In this work, if two branches do not share a common end, they are considered as electrically distant for simplicity.

\textbf{T1:} no two branch in $a$ share a common end, i.e, $i \not \in B(j)$ for any $i,j\in a$.

\noindent\textbf{Loop constraint:} 
Meshed structures should be avoided in the distribution system. Thus, a restoration action should not be performed if it creates a loop of energized branches connected to the same source. 


\textbf{T2:} no state $t\in Post(s,a)$ represents a loop of energized branches connected to the same source. 


\textbf{DER (Distributed Energy Resources) constraints:} A DER can only feed a limited number of buses due to its capacity. Consequently, a restoration action connecting a bus to a DER can not be applied if the total consumption of the buses connected to the DER exceeds its capacity. Here, the capacity of each DER and the load at each bus are assumed to be known. The generation at $i^{th}$ DER is denoted as $P_{G,i}$ and the load at $i^{th}$ bus is denoted as $P_{L,i}$. Given a state $s = [s^1,\ldots, s^L] \in S$ and index of a source $k \leq M$, $ActBus(s,k)$ denotes the set of energized buses connected to the source $k$:
\begin{align}
    ActBus(s,k) = \{ i \in I_N \mid & s^j = E_k \text{ for a branch } j \\
    & \text{connected to bus } i \} \nonumber
\end{align}

 \textbf{E1:} for each $t\in Post(s,a)$ and DER $k \in I_M$, it holds that
\begin{equation}
    \left ( \sum_{i\in ActBus(t,k)} P_{L,i} \right ) \leq P_{G,k}
\end{equation}

\begin{remark}
If there is an energized path between two DERs $k,m \in I_M$, then they are treated as one with capacity $P_{G,k} + P_{G,m}$. In addition, the capacity of the transmission grid is assumed to be infinite; and thus, \textbf{E1} is not checked for $k=0$.
\end{remark}




\par \textbf{Voltage limitation constraints:} 
The voltage of a bus must be in permissible limits standardized between 0.95 \textit{pu} and 1.05 \textit{pu}. To guarantee that the voltage constraints are not violated during the restoration,  FBPF analysis is utilized during the system construction. An action $a$ is marked as infeasible at state $s$ if the voltage constraint for a bus is violated in a state $t \in Post(s,a)$. Given the system topology, and the set of buses energized from a source, the FBPF analysis gives the set of buses at which the voltage constraints are violated. 

\textbf{E2:} for each $t\in Post(s,a)$ and source $k \in \{0\} \cup I_M$, it holds that
\begin{equation}
| FBPF( ActBus(s,k) ) | = 0
\end{equation}

\begin{remark}
For a state $s$, if no subset of $\bar A(s)$ satisfies all of the constraints, but some subsets of $\bar A(s)$ satisfies \textbf{T1}, \textbf{T2}, and \textbf{E1} (thus only violate \textbf{E2}), then the permissible limits are relaxed and the FBPF analysis is re-run. Note that the relaxation of the voltage limits are limited with the power quality standards.
\end{remark}





The summarized construction method guarantees that all system properties and constraints are integrated to the MDP model through the admissible control sets $A(s), s\in S$ and the transition function. Essentially, the construction guarantees that the restoration actions from set $a\in A(s)$ can be applied simultaneously when the configuration of the distribution system is $s$ and the restoration actions $a \notin A(s)$ violates a topological or electrical constraints of the system. 

\begin{figure}[ht]
\centering
\vspace{-12pt}
\includegraphics[scale=0.8]{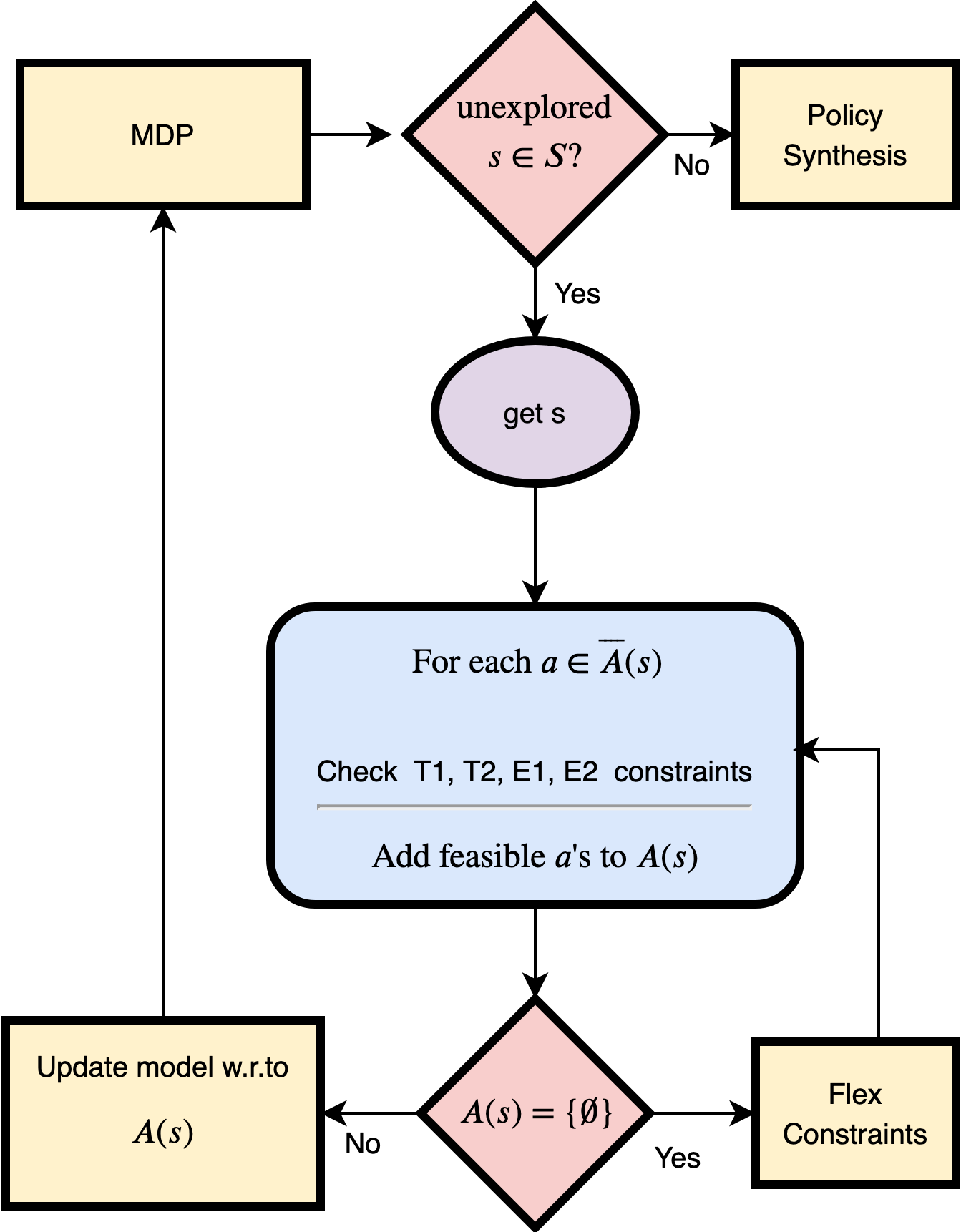}
\caption{MDP construction flowchart} \label{fig:flowchart}
\vspace{-12pt}
\end{figure}

The developed iterative MDP construction method is summarized in Fig.~\ref{fig:flowchart}. Initially, the set of unexplored states is $\{[U,\ldots,U]\}$, e.g., includes only the initial state. Then, iteratively, a state $s$ is taken from this set, the set of branches $\bar A(s)$, that a restoration action can be applied, is computed as in~\eqref{eq:actionUconstraint}. Next, the subsets $a$ of $\bar A(s)$ for which the constraints \textbf{T1}, \textbf{T2}, \textbf{E1} and \textbf{E2} are satisfied are added to $A(s)$. If $A(s)$ only includes the empty set, $a=\{\}$, the permissible limits for the FBPF analysis is relaxed and the constraint check is repeated. Otherwise, for each $a\in A(s)$, $p(\cdot \mid s, a)$~\eqref{eq:transition_probability} and $Post(s,a)$~\eqref{eq:post} are computed, and the new states are added to the set of unexplored states. The iterative process continues until all of the reachable states are explored, which indicates that the model construction is completed.

\subsection{Cost Formulation}
In this work, the goal is to minimize the average restoration time for each branch. To achieve this task, the state cost is defined as the number of unenergized buses: 
\begin{equation}\label{eq:statecost}
    c(s) = N - \left | \bigcup_{k=0}^L ActBus(s,k) \right |
\end{equation}

Thus, for a policy $\pi$ the n-step cost~\eqref{eq:erl} of a state $s$ is the expected total number of unenergized buses over n-steps. For a state $s$, $v^n_{\pi}(s)/N$ is the average expected time to energize a bus. Considering that $N$ is constant, the optimal policy $\pi^\star$ is the one that minimizes the average restoration time over the buses for the considered horizon $n$. Here, the number of branches $L$ is used as the optimization horizon since the restoration process is always completed in $L$ steps, e.g., single source radial branches. 
\begin{figure}[ht]
\centering
\tikzset{every picture/.style={line width=0.75pt}} 

\begin{tikzpicture}[x=0.75pt,y=0.75pt,yscale=-1,xscale=1]

\draw    (181,135) -- (213,135) ;

\draw  [fill={rgb, 255:red, 0; green, 0; blue, 0 }  ,fill opacity=1 ] (213,135) .. controls (213,130.58) and (216.58,127) .. (221,127) .. controls (225.42,127) and (229,130.58) .. (229,135) .. controls (229,139.42) and (225.42,143) .. (221,143) .. controls (216.58,143) and (213,139.42) .. (213,135) -- cycle ;
\draw    (221,135) -- (264.22,112.91) ;
\draw [shift={(266,112)}, rotate = 512.9300000000001] [color={rgb, 255:red, 0; green, 0; blue, 0 }  ][line width=0.75]    (10.93,-3.29) .. controls (6.95,-1.4) and (3.31,-0.3) .. (0,0) .. controls (3.31,0.3) and (6.95,1.4) .. (10.93,3.29)   ;

\draw    (221,135) -- (269.12,152.32) ;
\draw [shift={(271,153)}, rotate = 199.8] [color={rgb, 255:red, 0; green, 0; blue, 0 }  ][line width=0.75]    (10.93,-3.29) .. controls (6.95,-1.4) and (3.31,-0.3) .. (0,0) .. controls (3.31,0.3) and (6.95,1.4) .. (10.93,3.29)   ;

\draw (121,136) node  [align=left] {{\small $s_7 =[E_0,D,U,U,E_1]$}};
\draw (332,149) node  [align=left] {{\small $s_9 =[E_0,D,D,U,E_1]$}};
\draw (332,112) node  [align=left] {{\small $s_8 =[E_0,D,E_0,U,E_1]$}};
\draw (218,118) node  [align=left] {\{3\}};
\end{tikzpicture}
\vspace{-8pt}
\caption{$Post(s_7,\{3\})$ for $s_7$ } \label{fig:transition2}
\vspace{-4pt}
\end{figure}
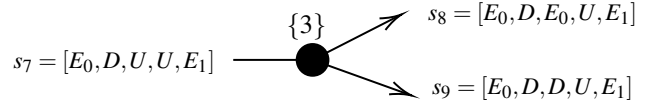

\begin{example}
For the system shown in Fig.~\ref{fig:sample}, consider $s_1 = [E_0,U,U,U,E_1]$ and $s_7$ shown in Fig.~\ref{fig:transition2}. Note that $Post(s_1, 2) = \{s_2, s_7\}$. 
The costs for the states from Figs.~\ref{fig:transition}-\ref{fig:transition2} are $c(s_1) = c(s_7) = c(s_9) = 2$, $c(s_2) = c(s_5) = c(s_6) =c(s_8)= 1$, $c(s_3) = c(s_4) = 0$. Consider a policy with $\pi(s_1) = \{2\}$,  $\pi(s_2) = \{3,4\}$, and $\pi(s_7) = \{3\}$. Assume that $P_f(i) = 0.4$ for each $i\in I_L$. The value function over two steps for $s_2$ and $s_7$ are $v^2_{\pi}(s_2) = 1.4$ and $v^2_{\pi}(s_7) = 3.4$, and the value function over three steps for $s_1$ is $v^3_{\pi}(s_1) = 4.2$. Given that the number of unenergized buses is $2$ at $s_1$, the average expected time to energize these branches is $2.1$ over a 3 step horizon. 
\end{example}

As illustrated in the example, the optimal solution for the finite horizon MDP policy synthesis problem gives the policy that minimize the average restoration time over the considered horizon thanks to the proposed state cost function~\eqref{eq:statecost}. Thus, the optimal policy gives the control action towards the states with higher number of energized buses.

\section{Case Study}
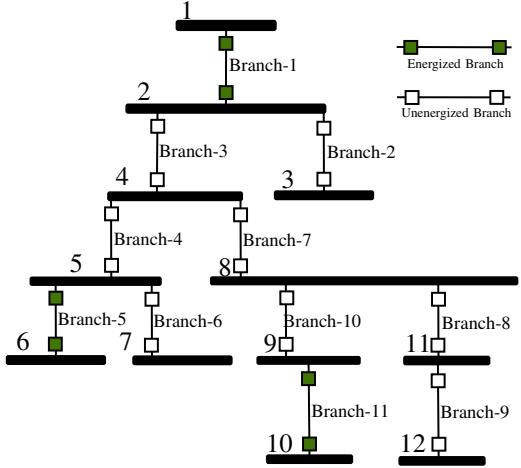
\begin{figure}[ht]
\centering

\tikzset{every picture/.style={line width=0.75pt}} 

\begin{tikzpicture}[x=0.75pt,y=0.75pt,yscale=-0.85,xscale=0.85]

\draw  [fill={rgb, 255:red, 0; green, 0; blue, 0 }  ,fill opacity=1 ][line width=0.75]  (357.83,15.59) .. controls (357.83,14.98) and (358.32,14.49) .. (358.93,14.49) -- (415.53,14.49) .. controls (416.13,14.49) and (416.63,14.98) .. (416.63,15.59) -- (416.63,18.89) .. controls (416.63,19.5) and (416.13,19.99) .. (415.53,19.99) -- (358.93,19.99) .. controls (358.32,19.99) and (357.83,19.5) .. (357.83,18.89) -- cycle ;
\draw    (387.23,17.24) -- (387.07,66.64) ;

\draw  [fill={rgb, 255:red, 0; green, 0; blue, 0 }  ,fill opacity=1 ][line width=0.75]  (328.04,65.16) .. controls (328.04,64.62) and (328.48,64.18) .. (329.02,64.18) -- (445.11,64.18) .. controls (445.65,64.18) and (446.1,64.62) .. (446.1,65.16) -- (446.1,68.11) .. controls (446.1,68.65) and (445.65,69.1) .. (445.11,69.1) -- (329.02,69.1) .. controls (328.48,69.1) and (328.04,68.65) .. (328.04,68.11) -- cycle ;
\draw  [fill={rgb, 255:red, 65; green, 117; blue, 5 }  ,fill opacity=1 ] (383.29,24.05) -- (390.95,24.05) -- (390.95,31.71) -- (383.29,31.71) -- cycle ;
\draw  [fill={rgb, 255:red, 65; green, 117; blue, 5 }  ,fill opacity=1 ] (383.38,53.29) -- (391.05,53.29) -- (391.05,60.95) -- (383.38,60.95) -- cycle ;
\draw  [fill={rgb, 255:red, 0; green, 0; blue, 0 }  ,fill opacity=1 ] (317.22,117.28) .. controls (317.22,116.77) and (317.63,116.35) .. (318.15,116.35) -- (395.74,116.35) .. controls (396.25,116.35) and (396.67,116.77) .. (396.67,117.28) -- (396.67,120.07) .. controls (396.67,120.58) and (396.25,121) .. (395.74,121) -- (318.15,121) .. controls (317.63,121) and (317.22,120.58) .. (317.22,120.07) -- cycle ;
\draw  [fill={rgb, 255:red, 0; green, 0; blue, 0 }  ,fill opacity=1 ] (416.1,116.68) .. controls (416.1,116.16) and (416.52,115.74) .. (417.04,115.74) -- (473.51,115.74) .. controls (474.03,115.74) and (474.46,116.16) .. (474.46,116.68) -- (474.46,119.51) .. controls (474.46,120.03) and (474.03,120.45) .. (473.51,120.45) -- (417.04,120.45) .. controls (416.52,120.45) and (416.1,120.03) .. (416.1,119.51) -- cycle ;
\draw [color={rgb, 255:red, 0; green, 0; blue, 0 }  ,draw opacity=1 ]   (346.5,67.8) -- (346.12,117.1) ;

\draw    (445.11,69.1) -- (445.28,118.1) ;

\draw  [fill={rgb, 255:red, 0; green, 0; blue, 0 }  ,fill opacity=1 ] (271.17,167.8) .. controls (271.17,167.31) and (271.56,166.92) .. (272.05,166.92) -- (347.55,166.92) .. controls (348.03,166.92) and (348.43,167.31) .. (348.43,167.8) -- (348.43,170.45) .. controls (348.43,170.94) and (348.03,171.33) .. (347.55,171.33) -- (272.05,171.33) .. controls (271.56,171.33) and (271.17,170.94) .. (271.17,170.45) -- cycle ;
\draw [color={rgb, 255:red, 0; green, 0; blue, 0 }  ,draw opacity=1 ]   (395.74,121) -- (395.9,170) ;

\draw  [fill={rgb, 255:red, 0; green, 0; blue, 0 }  ,fill opacity=1 ] (378.22,167.62) .. controls (378.22,167.16) and (378.59,166.78) .. (379.06,166.78) -- (559.16,166.78) .. controls (559.62,166.78) and (560,167.16) .. (560,167.62) -- (560,170.16) .. controls (560,170.62) and (559.62,171) .. (559.16,171) -- (379.06,171) .. controls (378.59,171) and (378.22,170.62) .. (378.22,170.16) -- cycle ;
\draw    (318.92,120.67) -- (319.08,169.67) ;

\draw  [fill={rgb, 255:red, 255; green, 255; blue, 255 }  ,fill opacity=1 ] (342.81,73.29) -- (350.48,73.29) -- (350.48,80.95) -- (342.81,80.95) -- cycle ;
\draw  [fill={rgb, 255:red, 255; green, 255; blue, 255 }  ,fill opacity=1 ] (342.52,105) -- (350.19,105) -- (350.19,112.67) -- (342.52,112.67) -- cycle ;
\draw  [fill={rgb, 255:red, 255; green, 255; blue, 255 }  ,fill opacity=1 ] (441.38,74.43) -- (449.05,74.43) -- (449.05,82.1) -- (441.38,82.1) -- cycle ;
\draw  [fill={rgb, 255:red, 255; green, 255; blue, 255 }  ,fill opacity=1 ] (441.38,104.71) -- (449.05,104.71) -- (449.05,112.38) -- (441.38,112.38) -- cycle ;
\draw  [fill={rgb, 255:red, 255; green, 255; blue, 255 }  ,fill opacity=1 ] (315.38,125.29) -- (323.05,125.29) -- (323.05,132.95) -- (315.38,132.95) -- cycle ;
\draw  [fill={rgb, 255:red, 255; green, 255; blue, 255 }  ,fill opacity=1 ] (315.38,155.86) -- (323.05,155.86) -- (323.05,163.52) -- (315.38,163.52) -- cycle ;
\draw  [fill={rgb, 255:red, 255; green, 255; blue, 255 }  ,fill opacity=1 ] (391.95,125.86) -- (399.62,125.86) -- (399.62,133.52) -- (391.95,133.52) -- cycle ;
\draw  [fill={rgb, 255:red, 255; green, 255; blue, 255 }  ,fill opacity=1 ] (391.95,156.14) -- (399.62,156.14) -- (399.62,163.81) -- (391.95,163.81) -- cycle ;
\draw    (488.39,30.02) -- (556.57,30.24) ;

\draw    (488.61,59.87) -- (556,59.95) ;

\draw  [fill={rgb, 255:red, 65; green, 117; blue, 5 }  ,fill opacity=1 ] (492.94,26.45) -- (500.61,26.45) -- (500.61,34.12) -- (492.94,34.12) -- cycle ;
\draw  [fill={rgb, 255:red, 65; green, 117; blue, 5 }  ,fill opacity=1 ] (545.48,26.49) -- (553.14,26.49) -- (553.14,34.15) -- (545.48,34.15) -- cycle ;
\draw  [fill={rgb, 255:red, 255; green, 255; blue, 255 }  ,fill opacity=1 ] (493.69,56.2) -- (501.36,56.2) -- (501.36,63.87) -- (493.69,63.87) -- cycle ;
\draw  [fill={rgb, 255:red, 255; green, 255; blue, 255 }  ,fill opacity=1 ] (543.98,56.13) -- (551.64,56.13) -- (551.64,63.8) -- (543.98,63.8) -- cycle ;
\draw  [fill={rgb, 255:red, 0; green, 0; blue, 0 }  ,fill opacity=1 ] (257.04,214.5) .. controls (257.04,213.98) and (257.46,213.56) .. (257.98,213.56) -- (314.45,213.56) .. controls (314.97,213.56) and (315.39,213.98) .. (315.39,214.5) -- (315.39,217.33) .. controls (315.39,217.85) and (314.97,218.27) .. (314.45,218.27) -- (257.98,218.27) .. controls (257.46,218.27) and (257.04,217.85) .. (257.04,217.33) -- cycle ;
\draw    (286.05,166.92) -- (286.22,215.92) ;

\draw  [fill={rgb, 255:red, 65; green, 117; blue, 5 }  ,fill opacity=1 ] (282.38,175.33) -- (290.05,175.33) -- (290.05,183) -- (282.38,183) -- cycle ;
\draw  [fill={rgb, 255:red, 65; green, 117; blue, 5 }  ,fill opacity=1 ] (282.05,202.62) -- (289.71,202.62) -- (289.71,210.29) -- (282.05,210.29) -- cycle ;
\draw  [fill={rgb, 255:red, 0; green, 0; blue, 0 }  ,fill opacity=1 ] (332.1,214.68) .. controls (332.1,214.16) and (332.52,213.74) .. (333.04,213.74) -- (389.51,213.74) .. controls (390.03,213.74) and (390.46,214.16) .. (390.46,214.68) -- (390.46,217.51) .. controls (390.46,218.03) and (390.03,218.45) .. (389.51,218.45) -- (333.04,218.45) .. controls (332.52,218.45) and (332.1,218.03) .. (332.1,217.51) -- cycle ;
\draw    (343.05,167.58) -- (343.22,216.58) ;

\draw  [fill={rgb, 255:red, 255; green, 255; blue, 255 }  ,fill opacity=1 ] (339.38,176) -- (347.05,176) -- (347.05,183.67) -- (339.38,183.67) -- cycle ;
\draw  [fill={rgb, 255:red, 255; green, 255; blue, 255 }  ,fill opacity=1 ] (339.05,203.29) -- (346.71,203.29) -- (346.71,210.95) -- (339.05,210.95) -- cycle ;
\draw  [fill={rgb, 255:red, 0; green, 0; blue, 0 }  ,fill opacity=1 ] (405.77,214.9) .. controls (405.77,214.44) and (406.14,214.07) .. (406.59,214.07) -- (465.67,214.07) .. controls (466.13,214.07) and (466.5,214.44) .. (466.5,214.9) -- (466.5,217.37) .. controls (466.5,217.83) and (466.13,218.2) .. (465.67,218.2) -- (406.59,218.2) .. controls (406.14,218.2) and (405.77,217.83) .. (405.77,217.37) -- cycle ;
\draw    (436.13,216.14) -- (436.7,274.8) ;

\draw  [fill={rgb, 255:red, 65; green, 117; blue, 5 }  ,fill opacity=1 ] (432.25,223.15) -- (439.92,223.15) -- (439.92,230.82) -- (432.25,230.82) -- cycle ;
\draw  [fill={rgb, 255:red, 65; green, 117; blue, 5 }  ,fill opacity=1 ] (432.85,261.95) -- (440.52,261.95) -- (440.52,269.62) -- (432.85,269.62) -- cycle ;
\draw  [fill={rgb, 255:red, 0; green, 0; blue, 0 }  ,fill opacity=1 ] (492.83,214.83) .. controls (492.83,214.33) and (493.24,213.92) .. (493.75,213.92) -- (542.48,213.92) .. controls (542.99,213.92) and (543.4,214.33) .. (543.4,214.83) -- (543.4,217.58) .. controls (543.4,218.09) and (542.99,218.5) .. (542.48,218.5) -- (493.75,218.5) .. controls (493.24,218.5) and (492.83,218.09) .. (492.83,217.58) -- cycle ;
\draw    (513.05,167.58) -- (513.22,216.58) ;

\draw  [fill={rgb, 255:red, 255; green, 255; blue, 255 }  ,fill opacity=1 ] (509.38,176) -- (517.05,176) -- (517.05,183.67) -- (509.38,183.67) -- cycle ;
\draw  [fill={rgb, 255:red, 255; green, 255; blue, 255 }  ,fill opacity=1 ] (509.05,203.29) -- (516.71,203.29) -- (516.71,210.95) -- (509.05,210.95) -- cycle ;
\draw  [fill={rgb, 255:red, 0; green, 0; blue, 0 }  ,fill opacity=1 ] (490.03,274.43) .. controls (490.03,273.93) and (490.44,273.52) .. (490.95,273.52) -- (539.68,273.52) .. controls (540.19,273.52) and (540.6,273.93) .. (540.6,274.43) -- (540.6,277.18) .. controls (540.6,277.69) and (540.19,278.1) .. (539.68,278.1) -- (490.95,278.1) .. controls (490.44,278.1) and (490.03,277.69) .. (490.03,277.18) -- cycle ;
\draw    (512.85,215.98) -- (513.33,276.2) ;

\draw  [fill={rgb, 255:red, 255; green, 255; blue, 255 }  ,fill opacity=1 ] (509.18,224.4) -- (516.85,224.4) -- (516.85,232.07) -- (509.18,232.07) -- cycle ;
\draw  [fill={rgb, 255:red, 255; green, 255; blue, 255 }  ,fill opacity=1 ] (509.65,262.09) -- (517.31,262.09) -- (517.31,269.75) -- (509.65,269.75) -- cycle ;
\draw    (422.62,167.71) -- (422.79,216.71) ;

\draw  [fill={rgb, 255:red, 255; green, 255; blue, 255 }  ,fill opacity=1 ] (419.38,175.15) -- (427.05,175.15) -- (427.05,182.82) -- (419.38,182.82) -- cycle ;
\draw  [fill={rgb, 255:red, 255; green, 255; blue, 255 }  ,fill opacity=1 ] (418.98,202.75) -- (426.65,202.75) -- (426.65,210.42) -- (418.98,210.42) -- cycle ;
\draw  [fill={rgb, 255:red, 0; green, 0; blue, 0 }  ,fill opacity=1 ] (411.42,273.43) .. controls (411.42,272.92) and (411.83,272.51) .. (412.33,272.51) -- (461.07,272.51) .. controls (461.57,272.51) and (461.98,272.92) .. (461.98,273.43) -- (461.98,276.18) .. controls (461.98,276.68) and (461.57,277.09) .. (461.07,277.09) -- (412.33,277.09) .. controls (411.83,277.09) and (411.42,276.68) .. (411.42,276.18) -- cycle ;

\draw (364.13,8.09) node  [align=left] {1};
\draw (338.63,56.2) node  [align=left] {2};
\draw (325.67,106.67) node  [align=left] {4};
\draw (424,108.33) node  [align=left] {3};
\draw (386.67,160.67) node  [align=left] {8};
\draw (298.47,158.6) node  [align=left] {5};
\draw (409.5,40.5) node  [align=left] {{\scriptsize Branch-1}};
\draw (467.5,92.5) node [scale=0.7] [align=left] {Branch-2};
\draw (368,91.5) node [scale=0.7] [align=left] {Branch-3};
\draw (417.5,144.5) node [scale=0.7] [align=left] {Branch-7};
\draw (341,144) node [scale=0.7] [align=left] {Branch-4};
\draw (523.36,40.45) node [scale=0.5] [align=left] {Energized Branch};
\draw (523.93,69.6) node [scale=0.5] [align=left] {Unenergized Branch};
\draw (266.8,204.93) node  [align=left] {6};
\draw (327.8,204.6) node  [align=left] {7};
\draw (413.47,205.87) node  [align=left] {9};
\draw (418.8,265.47) node  [align=left] {10};
\draw (501.13,205.93) node  [align=left] {11};
\draw (497.93,265.13) node  [align=left] {12};
\draw (307.8,191.6) node [scale=0.7] [align=left] {Branch-5};
\draw (364.6,192.4) node [scale=0.7] [align=left] {Branch-6};
\draw (535.7,194.1) node [scale=0.7] [align=left] {Branch-8};
\draw (534.9,246.5) node [scale=0.7] [align=left] {Branch-9};
\draw (444.2,192.4) node [scale=0.7] [align=left] {Branch-10};
\draw (461,246) node [scale=0.7] [align=left] {Branch-11};

\end{tikzpicture}
\caption{12-bus test system} \label{fig:testSystem}
\vspace{-12pt}
\end{figure}

The proposed method is applied to a sample system shown in Fig.~\ref{fig:testSystem}. Bus 1 is connected to the transmission
grid, bus 6 and bus 10 are connected to DER-1 and DER-2, respectively. The power consumption at each bus is assumed to be the same, and the DER capacities are given as the number of buses that it can feed. 
The policy synthesis results for different DER capacities and different earthquake scenarios are presented in Table I. For each case, initially all circuit breakers are open and the branch statuses are unknown. 
\begin{table}[ht]
\scriptsize
\caption{System Properties of The Scenarios}
\label{tab:table1}
\centering

\begin{tabular}{|c|c|c|c|c|}
\hline 
       & DER Capacities         & \begin{tabular}[c]{@{}c@{}}Average Time\\ to Energize\end{tabular} & \begin{tabular}[c]{@{}c@{}}Number of\\  States\end{tabular} & \begin{tabular}[c]{@{}c@{}}$P_f$ value \\ for branches\end{tabular} \\ \hline
Case 1 & $DER_1$ = 1, $DER_2$ = 1 & 6.6 & 449 & 0.3 \\ \hline
Case 2 & $DER_1$ = 4, $DER_2$ = 4 & 5.2 & 1395 & 0.3  \\ \hline
Case 3 & $DER_1$ = 4, $DER_2$ = 4 & 2.77 & 1395 & 0.1   \\ \hline
Case 4 & $DER_1$ = 4, $DER_2$ = 4 & 3.39 & 1274 & various   \\ \hline

\end{tabular}
\end{table}

The size of the MDP model highly depends on the DER configuration as the number of the reachable states increases with the number of DERs  and their capacities. The effect of DER capacity on the MDP model is shown by Cases 1 and 2 given in Table~\ref{tab:table1}. In Case 1, the model has 449 states, while for Case 2 there exists 1274 states. For example in Case 1, when the system reaches state $[E_0,U,U,U,E_1,U,U,U,U,U,E_2]$, the restoration process can only continue from branch 1 as the DERs reached their capacity. On the other hand, in Case 2 for the same state, restoration process can continue from three different branches thanks to the increase in the DER capacities. The advantage of the DER capacity increase is the reduced expected bus restoration time as observed in Table~\ref{tab:table1}, since a DER with higher capacity can supply more customers, which enables an extended islanded operation until the connection with the transmission grid is established. 

The influence of $P_f$ can be observed when Case 2 and Case 3 are compared. Note that, as all the branches have the same $P_f$ in those cases, the synthesized restoration strategy will be the same for both of them. However, the average expected restoration time is reduced as the $P_f$ decreases. It can be concluded that the proposed cost formulation, therefore, provides the user an estimation over the damage status of the system via the average expected restoration time.

In Case 4, the $P_f$ of branch 4 is 0.7, and the rest of the $P_f$ values are $0.1$. The resulting restoration action sequences are shown below for cases 3 and 4. As seen, the method handles different $P_f$ values and generates a strategy accordingly. 
\begin{align*}
Case 3:&   \{1,5,11\} \rightarrow \{2,4,10\} \rightarrow \{3,6,8\} \rightarrow \{7,9\} \\
Case 4: & \{1,5,11\} \rightarrow \{2,6,10\} \rightarrow \{3,8\} \rightarrow \{4,9\} \rightarrow \{7\}
\end{align*}
In~\cite{AydinGol2019}, the objective was defined as minimizing the total restoration time. For Case 2, the average expected restoration time for the policy optimizing this objective is $5.52$. Essentially, minimizing the restoration time
make the policy choose the actions that likely end up in terminal states ($A(s) =\{\emptyset\}$) regardless of the energized buses.
Thus, it does not necessarily minimize the average expected restoration time of the buses, which is achieved with the new cost formulation.

The performance of the developed method is evaluated on the IEEE-34 bus system with different DER configurations (up to 2 DERs). The optimal policy is always synthesized within 10 minutes when $P_f(i) < 1$ for each branch. However, the computation time reduces significantly when some branches are known to be damaged, e.g. $P_f(i) = 1$.

\section{Conclusion}
An MDP based decision support method for earthquake affected distribution systems is developed in this paper. The proposed method employs $P_f$ values, which are calculated via the real time data recorded during the earthquake in order to determine probability of failure of the field components. The method minimizes the average expected restoration time of the buses, and provides a restoration strategy. During the modeling phase both the topological and electrical constraints are considered, which yields an admissible result.  
\ifCLASSINFOpdf


\section*{Acknowledgment}

This work is supported by Scientific and Technological Research Council of Turkey (TUBITAK) under project number 118E183, and the European Union’s Horizon 2020 research and innovation program under the Marie Sklodowska-Curie grant agreement No 798482.



\bibliographystyle{IEEEtran}
\bibliography{bibtex.bib}
%




\end{document}